\documentclass[12pt]{article}
\usepackage{amssymb}

\usepackage{cite} %makes references like [1-4] rather than [1,2,3,4]
\usepackage[english]{babel}
\usepackage{hyperref}
\usepackage{xcolor}
\usepackage{bbold} % Enables to use mathbb 1, but it also changes a bit all of the other boldfaces. If you do not like it remmeber to reinstate the command \BI

\newcommand{\be}{\begin{equation}}
\newcommand{\ee}{\end{equation}}
\newcommand{\bea}{\begin{array}}
\newcommand{\ea}{\end{array}}
\newcommand{\beqa}{\begin{eqnarray}}
\newcommand{\eeqa}{\end{eqnarray}}
\newcommand{\bean}{\begin{eqnarray*}}
\newcommand{\eean}{\end{eqnarray*}}

\def\up#1{\leavevmode \raise.16ex\hbox{#1}}

\newcommand{\gapproxeq}{\lower
.7ex\hbox{$\;\stackrel{\textstyle >}{\sim}\;$}}
\newcommand{\lapproxeq}{\lower .7ex\hbox{$\;\stackrel
{\textstyle <}{\sim}\;$}}
\newcounter{appendice}

\def\thebibliography#1{{\bf REFERENCES\markboth
{REFERENCES}{REFERENCES}}\list
{[\arabic{enumi}]}{\settowidth\labelwidth{[#1]}\leftmargin\labelwidth
\advance\leftmargin\labelsep
\usecounter{enumi}}
\def\newblock{\hskip .11em plus .33em minus -.07em}
\sloppy
\sfcode`\.=1000\relax}

\newcommand{\al}[1]{\textcolor{black}{#1}}

\begin{document}

\centerline{\large   Symmetry Nonrestoration  in the Pati–Salam Model
 }
\vskip .5cm

\centerline{ Nobuchika Okada\footnote{okadan@ua.edu} and Allen Stern\footnote{astern@ua.edu} }

\vskip 1cm

\begin{center}
  { Department of Physics, University of Alabama,\\ Tuscaloosa,
Alabama 35487, USA\\}

\end{center}

\bigskip
\bigskip

\abstract{We demonstrate that symmetry need not be restored in the Pati-Salam model, so that  $SU(2)_L\otimes SU(2)_R\otimes SU(4)_c$  remains broken to   the Standard Model group at temperatures above the Pati-Salam symmetry breaking scale. Including the leading finite-temperature corrections, suitable quartic couplings prevent a restoration transition,  thereby avoiding the thermal production of 't Hooft-Polyakov monopoles after inflation, even if the reheating temperature is very high. This removes monopole-based constraints on the Pati-Salam symmetry breaking scale.}

\newpage

Results from Planck 2018\cite{Planck:2018jri} and combined Planck+BICEP/Keck 2018 analysis, \cite{BICEP:2021xfz} can be used to estimate  an  upper bound for   the reheating temperature after inflation $T_R$.  This bound is obtained within the context of  
the standard slow-roll inflation scenario,  the derivation of which can be summarized as follows:
In the standard slow-roll inflation scenario, cosmic inflation is driven by a scalar field (the inflaton) as it slowly rolls down toward the minimum of its potential.
The amplitude of the curvature perturbation is given by
\begin{eqnarray}
\Delta_{\mathcal R}^2
= \frac{1}{24\pi^2}\,
\frac{V(\phi_k)}{M_P^4\,\epsilon(\phi_k)} ,
\label{pws}
\end{eqnarray}
where $V(\phi_k)$ is the inflaton potential evaluated at the inflaton field value $\phi_k$ when the pivot scale $k=0.05~\mathrm{Mpc}^{-1}$ exits the horizon.
The slow-roll parameter $\epsilon(\phi_k)$ is related to the observable tensor-to-scalar ratio via $r = 16\,\epsilon(\phi_k)$, and
$M_P = 2.4 \times 10^{18}~\mathrm{GeV}$ denotes the reduced Planck mass.
This prediction should be compared with the Planck 2018 result \cite{Planck:2018jri},
\begin{eqnarray}
\Delta_{\mathcal R}^2 = 2.1 \times 10^{-9},
\end{eqnarray}
while the combined Planck+BICEP/Keck 2018 analysis \cite{BICEP:2021xfz} yields the upper bound $r < 0.036 $ at 95\% Confidence Level. 
In certain classes of inflationary scenarios, reheating after inflation can occur very efficiently, for example through preheating \cite{Kofman:1994rk,Kofman:1997yn} 
 or violent preheating \cite{Ema:2016dny}. (For general discussions see, e.g., \cite{Bassett:2005,Lozanov:2019}.)
In such cases, assuming an almost instantaneous conversion of the inflaton energy into radiation, the reheating temperature $T_R$ is roughly determined by equating the inflaton energy density to that of a thermal plasma,
\begin{eqnarray}
V(\phi_k) \simeq \frac{\pi^2}{30}\, g_*\, T_R^4 ,
\end{eqnarray}
where $g_* \simeq 100$ is the effective number of relativistic degrees of freedom.
Using Eq.~(\ref{pws}) together with the Planck 2018 normalization, the reheating temperature can be expressed 
as a function of the tensor-to-scalar ratio $r$:
\begin{eqnarray}
T_R \,[\mathrm{GeV}] \simeq 1.3 \times 10^{16}\, r^{1/4}.
\end{eqnarray}
The current upper bound $r < 0.036$ therefore implies
%\begin{eqnarray}
$T_R \lesssim 5.8 \times 10^{15}~\mathrm{GeV}$,
%\end{eqnarray}
which is close to the typical grand unified theory (GUT) scale.
Future high-sensitivity cosmic microwave background (CMB) polarization experiments, 
   such as CMB-S4 \cite{Abazajian:2019eic} and their successors, and  the planned LiteBIRD satellite mission \cite{LiteBIRD:2022cnt}, 
   will probe tensor-to-scalar ratios down to $r = \mathcal{O}(10^{-3})$ at 95\% Confidence Level. 
This corresponds to reheating temperatures as low as
\begin{eqnarray}
T_R \simeq 2.4 \times 10^{15}~\mathrm{GeV}.
\end{eqnarray}   
If the Universe is reheated to a temperature higher than this value, the corresponding tensor-to-scalar ratio is within reach 
  of future CMB polarization measurements.

The reheating temperature after inflation $T_{R}$ plays a role in the construction of grand unified theories, since a gauge symmetry whose restoration temperature $T_{res}$ is below $T_{R}$ can be thermally restored   after inflation.  This potentially  leads  to an unwanted overproduction of magnetic monopoles, or other possible topological defects after cooling.  As a result, the symmetry restoration temperature $T_{res}$ for such theories is required to be greater than $T_{R}$.

 The bound on the symmetry restoration temperature is of particular interest for the Pati-Salam model,\cite{Pati:1974yy} and its supersymmetric extensions,   because it has been considered for a very wide range of energy scales.\cite{Antusch:2014}  
 't Hooft-Polyakov monopoles, as well as other exotic states, occur in this model due to the breaking of the Pati-Salam symmetry group $G_{PS}=SU(2)_L\otimes SU(2)_R\otimes SU(4)_c$ to the standard model subgroup $G_{SM}=SU(2)_L\otimes U(1)_Y\otimes SU(3)_c$.\cite{Kephart:2025tik}  The proposed Pati-Salam energy scales range  from  near GUT scales, $\sim 10^{15}-10^{16}$ GeV, to the intermediate scales, $\sim 10^{12}- 10^{14}$ GeV (the natural see-saw scale), all the way down to low energy scales, TeV-PeV.  At first sight, the lower bound on $T_{res}$ appears to place strong restrictions on intermediate energy Pati-Salam models, and to essentially rule out low scale Pati–Salam scenarios.  However, it is known that the symmetry restoration temperature need not correspond to the nominal symmetry breaking scale of a gauge theory.   Symmetry restoration may occur at a higher temperature than the symmetry breaking scale, and for a judicious choice of scalar field self couplings  the symmetry may not be restored at any temperature. This was shown   long ago, in particular, for the minimal $SU(5)$ GUT model,
 as one means of  eliminating the monopole overproduction in that model.\cite{Salomonson:1984rh,Dvali:1995cj} (See also \cite{Mohapatra:1979}-\cite{Bajc:1998}.) 
  In this article, we demonstrate that an analogous mechanism operates in the Pati–Salam model.
  We take the Higgs fields representations in the model to be   $(1,2,4)$,  $(1,2,\overline{4})$ and $(2,2,1)$ of $G_{PS}$, although the procedure should be straightforwardly  adaptable to other  suitable Higgs representations.  As in \cite{Salomonson:1984rh}, we shall work at tree level for the scalar potential, with leading ${\cal O}(T^2)$
 thermal corrections in the high‑temperature expansion. \al{Higher-order effects, including gauge, Yukawa, and daisy-resummed contributions, may modify the quantitative details of the parameter space, but are not expected to alter the qualitative possibility of symmetry non-restoration within the perturbative regime.} We show that for suitable choices of Higgs  quartic couplings, the Pati–Salam symmetry can remain broken at all temperatures, so that no monopole producing phase transition occurs after inflation. As a consequence, there need be no constraint on the Pati–Salam breaking scale arising from $T_{R}$, even in models with relatively low scale Pati–Salam unification.\footnote{An alternative proposal for surpressing monopole production in the Pati
Salam model was given in \cite{Senjanovic:2025enc}.}

The outline for this article is as follows: We will begin by first introducing our notation and then writing down the Higgs potential involved in the breaking of $G_{PS}$ to $G_{SM}$.  An analysis will be done to obtain various inequalities relating  the  various coefficients of the potential will be performed, and leading temperature corrections to the potential will  be included.

Our notation for the relevant groups and their embeddings, along with the representations of the Higgs fields now follows.
We denote the elements of  $G_{PS}$   by  $(u^L,u^R,g)_{PS}$, where $u^L,u^R,g$ are elements of  $SU(2)_L\,, \;  SU(2)_R$ and $ SU(4)_c$, respectively, in their defining representations,   \beqa && u^L=[u^L_{i'j'}]\in SU(2)_L\,,\quad\;\; i',j',...=1,2\cr  &&u^R=[u^R_{ij}]\in SU(2)_R\,, \qquad i,j,...=1,2\cr&& g=[g_{\alpha\beta}]\in SU(4)\,,\qquad  \;\alpha,\beta...=1,2,3,4\eeqa
 We denote the elements  of the standard model subgroup $G_{SM}$ by $( u^L,e^{i\chi},\tilde g)_{SM}$, where the phase  $e^{i\chi}$ belongs to $ U(1)_Y$ and $\tilde g=[\tilde g_{ab}],  \;a,b...=1,2,3$ belongs to $ SU(3)_c$ in the defining representation.   The embedding of $G_{SM}$ into $G_{PS}$  is
given by
\be ( u^L,e^{i\chi},\tilde g)_{SM} \longmapsto \Biggl(u^L,\pmatrix{e^{-3i\chi}&  \cr &e^{3i\chi} },\pmatrix{e^{i\chi}\tilde g&\cr&e^{-3i\chi}}\Biggr)_{PS} \;. \label{stbgrp2}\ee
Call the elements of the  low energy subgroup $G_{LE}=U(1)\otimes SU(3)_c$, associated with the final stage of the symmetry breaking,  by $( e^{i\phi},\tilde g)_{LE}$, where the $U(1)$ phase $e^{i\phi}$ corresponds to electromagnetism.  Its 
 embedding into $SU(2)_L\otimes SU(2)_R\otimes SU(4)$  is
given by
\be (e^{i\phi},\tilde g)_{LE} \longmapsto \Biggl(\pmatrix{e^{3i \phi}&  \cr &e^{-3i\phi} },\pmatrix{e^{-3i\phi}&  \cr &e^{3i\phi} },\pmatrix{e^{i\phi}\tilde g&\cr&e^{-3i\phi}}\Biggr)_{PS} \;. \label{lebgrp}\ee  
We denote 
the Higgs fields by $\Phi_{i\alpha}$, $\Xi_{i\alpha }$ and  $H_{i' i}$, associated respectively with the $(1,2,4)$,  $(1,2,\overline{4})$ and $(2,2,1)$ of $G_{PS}$.
They transform  according to
\be \Phi_{i\alpha}\rightarrow{ \Phi'}_{i\alpha}=u^R_{ij}g_{\alpha\beta} \Phi_{j\beta}\;,\quad \Xi_{i\alpha}\rightarrow{ \Xi'}_{i\alpha}=u^R_{ij}{ g}_{\alpha\beta}^* \Xi_{j\beta}\,,\quad H_{i' i}\rightarrow H'_{i' i}=u^L_{i'j'}u^R_{ij}H_{j' j}\;.\ee
The spontaneous symmetry breaking from $G_{PS}$ to $G_{SM}$ arises from  vacuum values of  $\Phi_{i\alpha}$ and $\Xi_{i\alpha }$, while  vacuum values of $H_{i' i}$ break $G_{SM}$ to  $G_{LE}$.  The energy scales involved in the breaking will be denoted by $v$ and $\epsilon$, respectively, with $v>>\epsilon$.

Next we write down the Higgs potential up to fourth order. It is required to be  invariant  under $G_{PS}$. Let us  first only consider the potential involved in the breaking of  $G_{PS}$ to $G_{SM}$.  Since  the  $H_{i' i}$ fields play no role at this stage of the symmetry breaking, we now just look at the $\Phi_{i\alpha}$ and $\Xi_{i\alpha }$ dependence of the potential.  Denoting the quadratic and quartic terms by $V_{PS}^{(2)}$ and $V_{PS}^{(4)}$, respectively, we have
\beqa V_{PS}(\Phi,\Xi) &=& V_{PS}^{(2)}(\Phi,\Xi)+ V_{PS}^{(4)}(\Phi,\Xi)\;.\label{totlpot}
\eeqa
The most general $G_{PS}$ invariant quadratic terms are
\beqa V_{PS}^{(2)}(\Phi,\Xi)&=&a_1 \Phi_{i\alpha}^*\Phi_{i\alpha}+a_2 \Xi_{i\alpha}^*\Xi_{i\alpha}+a_3\epsilon_{ij}\left(\Phi_{i\alpha}\Xi_{j\alpha}+c.c.\right)\;,\label{quadpot}
\eeqa $c.c.$ denoting complex conjugate terms.
Many $G_{PS}$ invariant quartic terms are possible.  We shall only consider a subset which is sufficient for our purposes:
\beqa V_{PS}^{(4)}(\Phi,\Xi)&=&a_4 \left(\Phi_{i\alpha}^*\Phi_{i\alpha}\right)^2+a_5 \left(\Xi_{i\alpha}^*\Xi_{i\alpha}\right)^2\
+a_6|\epsilon_{ij}\Phi_{i\alpha}\Xi_{j\alpha}|^2\cr&&\cr&&+a_7 \epsilon_{ij}\epsilon_{k\ell}\Phi_{i\alpha}^*\Phi_{k\alpha}\Xi_{j\beta}^*\Xi_{\ell\beta}
+ a_8\Phi_{i\alpha}^*\Phi_{i\beta}\Phi_{j\beta}^*\Phi_{j\alpha}\cr&&\cr&&+ a_9\Xi_{i\alpha}^*\Xi_{i\beta}\Xi_{j\beta}^*\Xi_{j\alpha}+a_{10}\Phi_{i\alpha}^*\Xi_{i\beta}\Xi_{j\alpha}^*\Phi_{j\beta}\;.\label{qurticPS}
\eeqa
$a_{1-10}$ are real parameters, whose values shall be restricted by  various inequalities obtained by a detailed analysis which we carry out below.  $a_{10}$ will play an important role in the analysis.

The spontaneous symmetry breaking of $G_{PS}$ to $G_{SM}$ can be implemented using the following vacuum expectation values for  $\Phi_{i\alpha}$ and $\Xi_{i\alpha }$:
\be <\Phi_{i\alpha}>=  v\delta_{i2}\delta_{\alpha 4}\;,\qquad\al{<\Xi_{i\alpha}>=0} \;.\label{phivevs}\ee 
We take  $v$ to be real. \al{We could also give a nonvanishing vacuum expectation value for $\Xi_{14}$ but this is not necessary at this stage of the symmetry breaking}.   Next let's perturb $\Phi_{i\alpha }$ and $\Xi_{i\alpha }$ about their vacuum values
\be \Phi_{i\alpha }= v\delta_{i2}\delta_{\alpha 4}+\phi_{i\alpha }\;,\qquad   \al{\Xi_{i\alpha }= \xi_{i\alpha }}\;.
\label{pertPhi}   \ee
Upon substituting into the Higgs potential (\ref{totlpot}), and collecting all the linear terms in  
$\phi_{i\alpha }$ and $\xi_{i\alpha }$  we get:\al{
\be\left(a_1v+2a_4 v^3+2a_8v^3\right)\left(\phi_{24}+\phi_{24}^*\right)  
- a_3 v\left(\xi_{14}+\xi_{14}^*\right) \label{pslin}\;.\ee}
They need to  vanish, which leads to
\al{\be a_1+2(a_4 +a_8)v^2=a_3=0 \;. \label{frsorcon}\ee}
After applying these conditions, the quadratic terms in $\phi_{i\alpha}$ and  $\xi_{i\alpha}$  are
\al{\beqa&&  a_2\xi^*_{i\alpha}\xi_{i\alpha}\,+\,v^2\biggl(
a_4 \left(\phi_{2\,4}+\phi_{2\,4}^*\right)^2+a_6|\xi_{14}|^2
+
a_7\xi_{1\alpha}^*\xi_{1\alpha}+a_{10}|\xi_{24}|^2\cr&&\cr&&+a_8 \left(2\phi_{i4}^*\phi_{i4}+2\phi_{2\alpha}^*\phi_{2\alpha}+(\phi_{24}^*)^2+(\phi_{24})^2-2\phi_{i\alpha}^*\phi_{i\alpha}\right)
\biggl)\;.\label{masterms}
\eeqa  Note that the coefficients $a_1$ and $a_3$ were eliminated by (\ref{frsorcon}), while the the terms associated with $a_5$ and $a_9$ don't contribute at second order.  The mass matrix in terms of real scalars is diagonal, with the  nonvanishing diagonal matrix  elements equal to }
\al{\be -2 a_8 v^2\;,\quad 4(a_4+a_8)v^2 \;,\quad a_2+a_7v^2\;,\quad a_2+(a_6+a_7)v^2\;,\quad a_2\;,\quad a_2+a_{10}v^2\;.\label{poseigvls}\ee
Local stability then demands  that all of these eigenvalues are positive, and thus
\be 0>a_8>-a_4\;,\qquad a_2>0,-a_{10}v^2,-a_7 v^2,-(a_6+a_7 )v^2\;.\label{nineteen}\ee 
From (\ref{frsorcon}) this also implies that 
\be  a_1<0\;.\label{twenty}\ee In addition to the scalar fields associated with  masses  (\ref{poseigvls}), there are nine zero modes:
\be \phi_{14},\phi_{21},\phi_{22},\phi_{23},\Im\phi_{24}\;.\label{gbsfststg}\ee
They  account for all of the Goldstone bosons, as there are nine broken symmetry generators at this stage of the symmetry breaking.
}

\al{The above considerations do not take into account the two quartic terms in $\Xi$ appearing in (\ref{qurticPS}), associated with the coefficients $a_5$ and $a_9$.  Since the former is positive definite, a necessary condition in order to ensure a lower bound on the potential is
\be a_5>0\;.\label{twntee2}\ee  Moreover, $a_5$ must be sufficiently larger than $|a_9|$ such that their combined  contribution to   (\ref{qurticPS}) is positive  as $\Xi\rightarrow\infty$ for all directions in target space.}

The leading temperature corrections to the Higgs potential go like $\frac {T^2}{24}$ times the trace of the second derivatives of the potential with respect to the Higgs fields.\cite{Dolan:1973qd}  In computing the second derivatives of $V_{PS}(\Phi,\Xi)$ with respect to $\Phi_{i\alpha}$ and $\Xi_{i\alpha}$, we can ignore the contributions from $V^{(2)}_{PS}$ since they will only shift the origin of the potential.
The trace is the sum of  the following two contributions:
\beqa \frac{\partial^2}{\partial\Phi_{i\alpha}\partial\Phi_{i\alpha}^*} V^{(4)}_{PS}(\Phi,\Xi)&=&\left(18a_4+12 a_8\right)\Phi_{j\beta}^*\Phi_{j\beta}+\left(a_6+4a_7+a_{10}\right)
 \Xi_{\ell\gamma}^*\Xi_{\ell\gamma}\qquad\qquad
\cr&&\cr
 \frac{\partial^2}{\partial\Xi_{i\alpha}\partial\Xi_{i\alpha}^*} V^{(4)}_{PS}(\Phi,\Xi)&=&\left(18a_5+12 a_9\right)\Xi_{j\beta}^*\Xi_{j\beta}+\left(a_6+4a_7+a_{10}\right)
 \Phi_{\ell\gamma}^*\Phi_{\ell\gamma}\;.\qquad\qquad\eeqa
So the leading temperature dependent corrections to the quadratic terms $V^{(2)}_{PS}$ in the potential are
 \be\frac {T^2}{24}\Bigl\{\left( 18a_4 +12a_8+a_6+4a_7 +a_{10}\right)\Phi_{j\beta}^*\Phi_{j\beta}
 +\left(18a_5+12a_9+a_6+4a_7 +a_{10}\right)   \Xi_{j\beta}^*\Xi_{j\beta}\Bigr\}\;.\ee
This means that the coefficients $a_1$ and $a_2$ effectively pick up the following temperature dependent corrections
\beqa a_1&\rightarrow& \tilde a_{1}(T)=a_1+\frac {T^2}{24}\left(18a_4 +12a_8+a_6+4a_7 +a_{10}\right)\cr&&\cr
 a_2&\rightarrow&\tilde a_{2}(T)= a_2+\frac {T^2}{24}\left(18a_5+12a_9+a_6+4a_7 +a_{10}\right) \;,\label{effquads}
\eeqa
at leading order.  The conditions (\ref{nineteen}) and (\ref{twenty}), which are necessary for the existence of a broken symmetry vacuum state, now become
\al{\be  \tilde a_1(T)<0\;,\qquad \tilde a_2(T)>0,-a_{10}v^2,-a_7 v^2,-(a_6+a_7 )v^2\;.\ee 
 That is, the Pati-Salam symmetry remains spontaneously broken at  temperature $T$ provided that 
\beqa  \frac {T^2}{24}\left(18a_4 +12a_8+a_6+4a_7 +a_{10}\right)&<&-a_1\label{ineqone}\\&&\cr \frac {T^2}{24}\left(18a_5+12a_9+a_6+4a_7 +a_{10}\right)&>&-a_2,-a_2-a_{10}v^2,-a_2-a_7 v^2,-a_2-(a_6+a_7 )v^2\;.\label{ineqtwo}
\cr&&\eeqa
From (\ref{twenty}), we note that the right hand side of (\ref{ineqone}) is positive.  Concerning the left hand side, $a_8<0$ and $a_6+4a_7 +a_{10}$ can be chosen to be  arbitrarily negative within the perturbative regime (which can then affect the lower bound for $a_2$), and therefore the inequality  (\ref{ineqone}) can be satisfied for any temperature.  With regard to (\ref{ineqtwo}),  we found no upper bounds on the coefficients $a_5$ and $a_9$.  They therefore can be chosen to be arbitrarily positive  within the perturbative regime, and therefore the inequality  (\ref{ineqtwo}) can be satisfied for any temperature. Thus it is possible to simultaneously satisfy (\ref{ineqone}) and   (\ref{ineqtwo}) such that  
{\it the Pati-Salam symmetry is not  restored  at  $T_R\gg v$.}}

In summary, starting from the tree‑level potential 
for the  Higgs fields (\ref{totlpot}), local stability of the vacuum that breaks $G_{PS}$ to $G_{SM}$
  at zero temperature is guaranteed if the quartic couplings satisfy conditions (\ref{frsorcon}), (\ref{nineteen}), (\ref{twenty}) and (\ref{twntee2}).
Including leading thermal corrections, the effective quadratic coefficients $a_1$ and $a_2$ are replaced by  $\tilde a_1(T)$ and $\tilde a_2(T)$ in (\ref{effquads}),
and the broken‑phase vacuum persists at temperature 
$T$ if the conditions  (\ref{ineqone}) and   (\ref{ineqtwo})  hold.
If $a_6+4a_7 +a_{10}$ is chosen to be  sufficiently negative with $a_5$ and $a_9$ chosen to be sufficiently positive within the perturbative regime, symmetry breaking is present at all temperatures  ensuring that  $G_{PS}$
  is never thermally restored and no monopole‑producing phase transition occurs after inflation.

For completeness we  include the $(2,2,1)$ Higgs  $H_{i' i}$ couplings in the potential, as they are required for spontaneous symmetry breaking from $G_{SM}$ to $G_{LE}$.  They are associated with an energy scale $\epsilon \ll v$, and thus generally can only produce  tiny modifications to the previous results.  The new  $G_{PS}$ invariant quadratic terms are
\beqa V_{SM}^{(2)}(H)&=& b_1H_{i' i}^*H_{i' i}+b_2\left(\det{H}+c.c.\right)\;.
\eeqa 
For the new quartic terms we can take
\newpage
\beqa V_{SM}^{(4)}(\Phi,\Xi,H)&=&b_3\left(H_{i' i}^*H_{i' i}\right)^2+b_4\,|\det{H}|^2
\cr&&\cr&&+b_5\,\Phi_{i\alpha}^*\Phi_{j\alpha}H_{i'j}^*
H_{i' i}+b_6\,\Xi_{i\alpha}^*\Xi_{j\alpha}H_{i'j}^*
H_{i' i}\label{VSM4}\;.
\eeqa
The total potential 
 \beqa V(\Phi,\Xi,H)&=&V_{PS}(\Phi,\Xi)+V_{SM}^{(2)}(H)+V_{SM}^{(4)}(\Phi,\Xi,H)\;.\label{VSM}\eeqa then contains $16$ parameters.
 Then the standard model group is spontaneously broken to $G_{LE}$  upon implementing the following vacuum expectation values for  $H_{i' i}$:
\al{\be < H_{i' i}>=\epsilon\delta_{i'2}\delta_{i2}\;.\label{PHvev}\ee  Note that $H_{21}$ is not invariant under the action of the $U(1)$ phase in (\ref{lebgrp}).  In addition to (\ref{PHvev}), the vacuum preserving electromagnetic symmetry requires that we  have the most general vacuum values for $\Phi_{i\alpha}$ and $\Xi_{i\alpha}$.  So we replace (\ref{phivevs}) with
\be <\Phi_{i\alpha}>=  v\delta_{i2}\delta_{\alpha 4}\;,\qquad<\Xi_{i\alpha}>=\epsilon'\delta_{i1}\delta_{\alpha 4} \;.\ee 
where $\epsilon,\epsilon' \ll v$. }
Let's  choose  $\epsilon$ and $\epsilon'$ to be real, and perturb all the scalar fields, including  $\Phi_{i\alpha}$ and $\Xi_{i\alpha }$, about their vacuum values
\be   H_{i' i}= \epsilon\delta_{i'2}\delta_{i2}+  h_{i' i}\qquad \Phi_{i\alpha }= v\delta_{i2}\delta_{\alpha 4}+\phi_{i\alpha }\;,\qquad   \al{\Xi_{i\alpha }=\epsilon'\delta_{i1}\delta_{\alpha 4}+ \xi_{i\alpha }}\;.\label{pertH}\ee
Now  upon substituting into the total potential (\ref{VSM}) we  get the linear terms
\al{$$(a_1v+2(a_4+a_8)v^3-a_3\epsilon'+(a_6+a_7)v\epsilon'^2+b_5v\epsilon^2)\left(\phi_{24}+ \phi_{24}^*\right)$$
$$+\,(-a_3v+a_2\epsilon'+(a_6+a_7)v^2\epsilon'+2(a_5+a_9)\epsilon'^3)(\xi_{14}+\xi_{14}^*)$$
\be +\,(b_1+2b_3 \epsilon^2+b_5v^2)\epsilon(h_{22}+h_{22}^*)  + b_2\epsilon (h_{11}+h_{11}^*)\;,\ee}
which are required to vanish.
 This means that  the  conditions  in (\ref{frsorcon}) are perturbed by $\epsilon$ and $\epsilon'$ dependent terms.
We also need 
\al{\be\;b_1+2b_3 \epsilon\,^2+b_5 v^2=0\;,\qquad b_2=0\;,\label{Hcoii}\ee}
in order for the linear terms in $h_{i'i}$ to vanish.
 After imposing (\ref{Hcoii}), the quadratic terms in $V_{SM}^{(2)}(H)+V_{SM}^{(4)}(\Phi,\Xi,H) $ are 
$$b_3\epsilon^2(h_{22}+h_{22}^*)^2+b_4\epsilon^2|h_{11}|^2+(\epsilon'^2b_6 -b_5v^2) h^*_{i'1}h_{i'1}+b_5\epsilon^2\phi_{2\alpha}^*\phi_{2\alpha}+b_6\epsilon^2\xi_{2\alpha}^*\xi_{2\alpha}$$
\be b_5v\epsilon\left(h_{22}\phi_{24}+h_{22}^*\phi_{24}^*+h_{2i}\phi_{i4}^*+h_{2i}^*\phi_{i4}\right)+b_6\epsilon\epsilon'(\xi_{24} h_{21}+\xi_{24}^* h_{21}^*)\label{fnlqudpt}\ee There are additional $\epsilon'$ dependent corrections to (\ref{masterms}).  As expected there are three massless Goldstone modes associated with the three broken generators of the Standard model.  They correspond to $h_{12}$ and $\Im h_{22}$.\footnote{The previous Goldstone modes (\ref{gbsfststg}) will in general be modified by $\epsilon$ and $\epsilon'$ dependent terms, and moreover, in order to protect them from acquiring a mass, the parameters $\epsilon$ and $\epsilon'$ cannot be kept independent.}  From   the third term in (\ref{fnlqudpt}), the hierarchy of Higgs masses requires  that $b_5= {\cal O}\left(\frac {\epsilon^2}{v^2}\right) \ll 1$.
Upon examining the eigenvalues, vacuum stability of the physical scalar modes requires that
\be -b_5 v^2 + b_4 \epsilon^2+b_6\epsilon'^2>0\;,\ee
while the condition  (\ref{nineteen}) picks up $\epsilon$ and $\epsilon'$ dependent corrections. \al{  In addition to the above conditions, we need to ensure that the potential  (\ref{VSM4}) has a lower bound.  Therefore
\be  b_3>0\;,\ee and that $b_3$ is sufficiently larger than -$b_4$ if $b_4<0$.}
  As the quartic terms  in (\ref{VSM4}) are small ${\cal O}\left(\frac {\epsilon^2}{v^2}\right) $ relative to those in (\ref{qurticPS}), they will not significantly affect the temperature corrections.

In conclusion, the results presented here show that symmetry nonrestoration can be implemented consistently in the Pati–Salam model with a minimal set of Higgs fields, namely those in the   $(1,2,4)$,  $(1,2,\overline{4})$ and $(2,2,1)$ representations of $G_{PS}$,  and a restricted set of quartic couplings.  The procedure used here  closely parallels the one demonstrating the nonrestoration of symmetry  for minimal $SU(5)$ GUTs.\cite{Salomonson:1984rh}  A set of  inequalities was obtained for the quartic parameters that ensured both local stability \al{  and a finite lower bound} of the standard model vacuum at zero temperature.  After including leading order temperature corrections, it was shown that
 for an appropriate choice of couplings,  the symmetry restoring phase transition does not occur, and for such a choice  ’t~Hooft–Polyakov monopoles associated with the breaking of $G_{PS}$
  are never thermally regenerated after inflation.  Thus monopole constraints on the Pati-Salam scale  are  removed in this case.

A number of extensions of this work are possible. One direction is to include the full set of allowed quartic invariants in the Higgs potential and to analyze  how higher‑order finite‑temperature corrections modify the nonrestoration conditions.  Nonperturbative effects,  effects from  the gauge sector and Yukawa couplings, alternative Higgs representations and supersymmetric and noncommutative extensions can also be considered.  This article suggests that the large freedom of the Higgs sector may allow for the nonrestoration of the Pati-Salam symmetry to persist for all such generalizations.
 
\bigskip
\bigskip
\noindent
{\bf Acknowledgement}

The work of N.O. is supported in part by the United States Department of Energy Grant Nos.~DE-SC0012447, DE-SC0023713, and DE-SC0026347.


\bigskip
\bigskip
\begin{thebibliography}{99}

%\cite{Planck:2018jri}
\bibitem{Planck:2018jri}
Y.~Akrami \textit{et al.} [Planck],
``Planck 2018 results. X. Constraints on inflation,''
Astron. Astrophys. \textbf{641}, A10 (2020).
%4270 citations counted in INSPIRE as of 19 Jan 2026

%\cite{BICEP:2021xfz}
\bibitem{BICEP:2021xfz}
P.~A.~R.~Ade \textit{et al.} [BICEP and Keck],
``Improved Constraints on Primordial Gravitational Waves using Planck, WMAP, and BICEP/Keck Observations through the 2018 Observing Season,''
Phys. Rev. Lett. \textbf{127}, no.15, 151301 (2021).
%1271 citations counted in INSPIRE as of 16 Jan 2026

%\cite{Kofman:1994rk}
\bibitem{Kofman:1994rk}
L.~Kofman, A.~D.~Linde and A.~A.~Starobinsky,
``Reheating after inflation,''
Phys. Rev. Lett. \textbf{73}, 3195-3198 (1994).
%2052 citations counted in INSPIRE as of 14 Jan 2026

%\cite{Kofman:1997yn}
\bibitem{Kofman:1997yn}
L.~Kofman, A.~D.~Linde and A.~A.~Starobinsky,
``Towards the theory of reheating after inflation,''
Phys. Rev. D \textbf{56}, 3258-3295 (1997).
%2242 citations counted in INSPIRE as of 14 Jan 2026

%\cite{Ema:2016dny}
\bibitem{Ema:2016dny}
Y.~Ema, R.~Jinno, K.~Mukaida and K.~Nakayama,
``Violent Preheating in Inflation with Nonminimal Coupling,''
JCAP \textbf{02}, 045 (2017).
%171 citations counted in INSPIRE as of 19 Jan 2026

\bibitem{Bassett:2005}
B.~A.~Bassett, S.~Tsujikawa and D.~Wands,
``Inflation Dynamics and Reheating,''
Rev.\ Mod.\ Phys.\ {\bf 78}, 537 (2006).



\bibitem{Lozanov:2019}
K.~D.~Lozanov,
``Lectures on Reheating After Inflation,''
JCAP {\bf 06}, 032 (2019).

%\cite{Abazajian:2019eic}
\bibitem{Abazajian:2019eic}
K.~Abazajian, G.~Addison, P.~Adshead, Z.~Ahmed, S.~W.~Allen, D.~Alonso, M.~Alvarez, A.~Anderson, K.~S.~Arnold and C.~Baccigalupi, \textit{et al.}
``CMB-S4 Science Case, Reference Design, and Project Plan,''
[arXiv:1907.04473 [astro-ph.IM]].
%907 citations counted in INSPIRE as of 14 Jan 2026

%\cite{LiteBIRD:2022cnt}
\bibitem{LiteBIRD:2022cnt}
E.~Allys \textit{et al.} [LiteBIRD],
``Probing Cosmic Inflation with the LiteBIRD Cosmic Microwave Background Polarization Survey,''
PTEP \textbf{2023}, no.4, 042F01 (2023).
%467 citations counted in INSPIRE as of 19 Jan 2026



\bibitem{Pati:1974yy}
J.~C.~Pati and A.~Salam,
``Lepton Number as the Fourth Color,''
Phys. Rev. D \textbf{10}, 275-289 (1974)
[erratum: Phys. Rev. D \textbf{11}, 703-703 (1975)] .


\bibitem{Antusch:2014}
S.~Antusch, S.~F.~King, C.~Luhn and M.~Spinrath,
``Multiple Scales in Pati--Salam Unification Models,''
Nucl.\ Phys.\ B {\bf 885}, 1 (2014).


\bibitem{Kephart:2025tik}
T.~W.~Kephart and Q.~Shafi,
``Magnetic Monopoles and Exotic States in $SU(4)_c \times SU(2)_L \times SU(2)_R$,''
Phys.\ Rev.\ D (2025).

\bibitem{Salomonson:1984rh}
P.~Salomonson, B.~S.~Skagerstam and A.~Stern,
``On the Primordial Monopole Problem in Grand Unified Theories,''
Phys. Lett. B \textbf{151}, 243-246 (1985).



\bibitem{Dvali:1995cj}
G.~R.~Dvali, A.~Melfo and G.~Senjanovic,
``Is There a monopole problem?,''
Phys. Rev. Lett. \textbf{75}, 4559-4562 (1995).


\bibitem{Mohapatra:1979}
R.~N.~Mohapatra and G.~Senjanovic,
``High Temperature Behavior of Gauge Theories,''
Phys.\ Rev.\ Lett.\ {\bf 42}, 1651 (1979).

\bibitem{Cline:1994}
J.~M.~Cline, K.~Kainulainen and K.~A.~Olive,
``On the Electroweak Phase Transition, Baryogenesis and Symmetry Nonrestoration,''
Phys.\ Rev.\ D {\bf 49}, 6394 (1994).

\bibitem{Boer:1995}
D.~Boer and J.~Ignatius,
``On Symmetry Nonrestoration at High Temperature,''
Phys.\ Lett.\ B {\bf 366}, 235 (1996).


\bibitem{Bajc:1998}
B.~Bajc and G.~Senjanovic,
``High Temperature Symmetry Nonrestoration and Supersymmetry,''
Nucl.\ Phys.\ B {\bf 526}, 129 (1998).

\bibitem{Senjanovic:2025enc}
G.~Senjanovi{\'c} and M.~Zantedeschi,
``Minimal Pati-Salam theory: From cosmic defects to gravitational waves and colliders,''
Phys. Rev. D \textbf{112}, no.5, 055018 (2025).

\bibitem{Dolan:1973qd}
L.~Dolan and R.~Jackiw,
``Symmetry behavior at finite temperature,''
Phys.\ Rev.\ D {\bf 9}, 3320 (1974). 

\end{thebibliography}
\end{document}